\documentclass{ws-p8-50x6-00}

\def\lsim{\raise0.3ex\hbox{$\;<$\kern-0.75em\raise-1.1ex\hbox{$\sim\;$}}}
\def\gsim{\raise0.3ex\hbox{$\;>$\kern-0.75em\raise-1.1ex\hbox{$\sim\;$}}}

\begin{document}

\title{Desperately Seeking The Standard Model}

%\begin{figure}[t]
%\begin{center}
%\epsfxsize=10pc 
%\epsfbox{cuerdas2.ps}
%\end{center} 
%\end{figure}

\author{Carlos Mu\~noz}

\address{Departamento de F\'{\i}sica
Te\'orica C-XI and Instituto de F\'{\i}sica
Te\'orica C-XVI,\\
Universidad Aut\'onoma de Madrid,
Cantoblanco, 28049 Madrid, Spain. \\
E-mail: carlos.munnoz@uam.es}

%\author{Elsie Tan, Jessie Tan and R. Sankaran}

%\address{World Scientific Publishing Co Ltd, 
%57 Shelton Street, Covent Garden, London WC2H 9HE, England\\
%E-mail: wspc@wspc.ox.uk}  


\maketitle

\abstracts{In the mid eighties string phenomenology started.
Since then, its main objective, the search
of the standard model, has not been accomplished yet.
In this talk, on the ocassion of the 2nd International Conference on
String Phenomenology in 2003,
I will review this crucial issue.}

%\vspace{-0.5cm}

%\rightline{FTUAM 03/29}
%\rightline{IFT-UAM/CSIC-03-49}
%\rightline{November 2003}

%\vspace{-0.5cm}

\section{Introduction}

Since this is the last talk of this meeting, 
%String Phenomenology 2003,
and everybody is already exhausted, following the suggestion of the
organizers I will try to give an `entertaining' talk 
about string phenomenology. To tell you the truth, I do not know whether
something concerning string phenomenology can be entertaining for an
audience.
In any case, please do not take too seriously all things that I am going
to say. Some of them are jokes!, or perhaps exaggerations.

The outline of the talk is very simple.
Basically, it is divided in two parts. 
The first one is very brief and 
%In the first one
I will give an optimistic view about string theory and
phenomenology. 
Following Dyson's analogy
between the quantum field theory and the 19th-century chemistry--both
explain {\it how} but not {\it why}--one could also establish an
analogy between atomic physics and string theory. Atomic physics
was needed to answer the question {\it why} in chemistry,
string theory is supposed to answer the question {\it why}
in the standard model of particle physics. 
I will 
%briefly 
review
this attempt in the second part of this 
talk. In this sense, in that part 
I will give a more realistic view of string phenomenology.
Perhaps, some of you in the audience will consider
this view slightly pessimistic. 
Let us see!

\begin{table}[t]
\caption{Chart of the fundamental particles (1953) listed
in the order of their mass.
%Experimental Data bearing on $\Gamma(K \to \pi \pi \gamma)$
%for the $K^0_S$, $K^0_L$ and $K^-$ mesons.
\label{1953}}
\begin{center}
\footnotesize
\begin{tabular}{|c|c|c|l|}
\hline
{photon} 
%&\raisebox{0pt}[13pt][7pt]{$\Gamma(\pi^- \pi^0)\; s^{-1}$} &
%\raisebox{0pt}[13pt][7pt]{$\Gamma(\pi^-\pi^0\gamma)\; s^{-1}$} &{}
\\
\hline
{graviton}
\\
\hline
{neutrino}
\\
\hline
{electron}
\\
\hline
{positron}
\\
\hline
{positive mu meson}
\\
\hline
{negative mu meson}
\\
\hline
{neutral pi meson}
\\
\hline
{positive pi meson}
\\
\hline
{negative pi meson}
\\
\hline
{zeta meson ?}
\\
\hline
{neutral V-particle}
\\
\hline
{tau meson}
\\
\hline
{kappa meson}
\\
\hline
{positive chi meson}
\\
\hline
{negative chi meson}
\\
\hline
{proton}
\\
\hline
{neutron}
\\
\hline
{neutral V-particle}
\\
\hline
{positive V-particle ?}
\\
\hline
\end{tabular}
\end{center}
\end{table}

\begin{table}[t]
\caption{Chart of the fundamental particles in 2003
(modification of Table~\ref{1953} taking into account the
new experimental results since 1953).
%listed
%in the order of their mass.
\label{2003}}
\begin{center}
\footnotesize
\begin{tabular}{|c|c|c|l|}
\hline
{photon} 
%&\raisebox{0pt}[13pt][7pt]{$\Gamma(\pi^- \pi^0)\; s^{-1}$} &
%\raisebox{0pt}[13pt][7pt]{$\Gamma(\pi^-\pi^0\gamma)\; s^{-1}$} &{}
\\
\hline
{graviton}
\\
\hline
{neutrino({\bf s})}
\\
\hline
{electron}
\\
\hline
{positron}
\\
\hline
{positive mu meson}
\\
\hline
{negative mu meson}
\\
\hline
{neutral pi meson  ({\it non elementary})}
\\
\hline
{positive pi meson ({\it non elementary})}
\\
\hline
{negative pi meson ({\it non elementary})}
\\
\hline
{zeta meson ? ({\it non elementary})}
\\
\hline
{neutral V-particle ({\it non elementary})}
\\
\hline
{(positive and negative) tau meson}
\\
\hline
{kappa meson ({\it non elementary})}
\\
\hline
{positive chi meson ({\it non elementary})}
\\
\hline
{negative chi meson ({\it non elementary})}
\\
\hline
{proton ({\it non elementary})}
\\
\hline
{neutron ({\it non elementary})}
\\
\hline
{neutral V-particle ({\it non elementary})}
\\
\hline
{positive V-particle ? ({\it non elementary})} 
\\
\hline
{\bf gluons}
\\
\hline
{\bf quarks + antiquarks}
\\
\hline
{\bf W$^{\pm}$, Z$^0$}
\\
\hline
{\bf Higgs ?}
\\
\hline
\end{tabular}
\end{center}
\end{table}

\section{Optimistic View}

Dyson, in an article written in 1953 \cite{Dyson},
drew the following analogy between the quantum field theory and
the
19th-century chemistry: 
`The latter described the properties of the chemical elements and their
interactions. How the elements behave; it did not try to explain why
a particular set of elements, each with its particular properties, exists.
To answer the question {\it why}, completely new sciences were needed:
atomic and nuclear physics. (...)
The quantum field theory treats elementary particles just as 19th-century
chemists treated the elements. The theory is in its nature descriptive and
not explanatory. It starts from the existence of a specified list of
elementary particles, with specified masses, spins, charges and specified
interactions with one another. All these data are put into the theory
at the beginning. The purpose of the theory is simply to deduce from this
information what will happen if particle $A$ is fired at particle $B$ with
a given velocity. We are not yet sure whether the theory will be able to 
fulfill even this modest purpose completely. Many technical difficulties
have still to be overcome. One of the difficulties is that we
do not yet have the complete list of elementary particles
(see Table~\ref{1953}).
%for the list of elementary particles as thought in
%1953). 
Nevertheless
the successes of the theory in describing experimental results have been
striking. It seems likely that the theory in something like its present
form will describe accurately a very wide range of possible experiments.
This is the most that we would wish to claim for it'.

Now, in 2003, as shown in Table~\ref{2003}, 
we do have the complete list of elementary particles (at 
least at energies below the electroweak scale, and given some 
uncertainties 
related to the Higgs sector)--the proton, pi meson, etc.
which appeared in the chart of fundamental particles in Dyson's presentation
should be regarded as an amusing historical anecdote--and
we do know that the theory fulfilling the modest purpose mentioned by 
Dyson is the {\it standard model} \cite{Glashow}. 

What the 19th-century chemistry did with the chemical elements,
the 20th-century standard model does with the elementary particles.
It describes how the elementary particles behave but does not try to 
explain why a particular set of elementary particles, each with
its particular properties, exists. To answer the question {\it why} it 
seems
that new sciences,
as atomic and nuclear physics in
the case of chemistry, are not needed, but 
just new theories. This is precisely one of the purposes of  
{\it string theory} (as a matter of fact, originally, the main purpouse of
the {\it modern} string theory was ``simply'' to unify all gauge interactions 
with gravity \cite{Scherk} in a consistent 
way \cite{Green}).

As is well known,
in string theory the elementary particles are not point-like objects but
extended, string-like objects. It is still surprising that this 
apparently 
small change allows us to answer fundamental questions that in the 
context
of the quantum field theory of point-like particles cannot even be posed.
%why do we live in four dimensions?,
For example: 
Why is the standard model gauge group
$SU(3)\times SU(2)_L\times U(1)_Y$?
Why are there three families of particles?
%Why is the pattern of quark and lepton masses so strange?
Why is the mass of the electron $m_e=0.5$ MeV? 
Why is the fine structure constant $\alpha=1/137$?

In this sense one can have the temptation of thinking that the
%that somebody in this century
%will write a similar comment to the one
%written by Dyson in his article about chemists  
comment about chemists that Dyson wrote in his
article
%mentioned in the introduction
--`Looking backward, 
it is now clear that 19th-century chemists were
right to concentrate on the {\it how} and to ignore the {\it why}.
They did not have the tools to begin to discuss intelligently
the reasons for the individualities of the {\it elements}. They had to
spend a hundred years building up a good quantitative descriptive
theory before they could go further. And the result of their labors, 
the
classical {\it science of chemistry}, was not destroyed or superseded by the 
later insight that {\it atomic physics} gave.'--will 
be written similarly by somebody in the future
about 20/21th--century physicists, substituting elements by {\it 
elementary 
particles}, science of chemistry by {\it standard model} and 
atomic 
physics by {\it string theory}.

Of course, let us hope that in this case the task 
will be accomplished before
a hundred years since the standard model started to be built.
Otherwise, many of the people  following this talk (including the
speaker) will be probably dead and
buried!

\section{Realistic (Pessimistic?) View}

What is string phenomenology?
A possible answer to this question is to say that
string phenomenology is the search of the standard model in string
theory.
Of course, this is not the only task,
but clearly to found the standard model is a necessary condition
in string phenomenology. It would be a little bit annoying to explain the 
big bang singularity using strings but not to be able to obtain
the standard model!
In this sense, is is fair to say that almost 20 years have gone by 
since string phenomenology started, and the standard model has not
been found yet.

Now, is this really a big problem?
Perhaps, in order to answer this question, we should get
some inspiration from the three great leaders shown in Fig.~\ref{azores}.
They have not found the weapons of mass destruction yet,
but they want us to believe 
%pretend to convince us 
that they will found them
in a few months. In the same way, we have not found the
standard model yet, but we want the people to believe that
we will find it in a few years.

\begin{figure}[t]
%\figurebox{20pc}{15pc}{azores2.ps} 
%\figurebox{1pc}{1pc}{azores2.ps}
% to have a box alone
\begin{center}
\epsfxsize=17pc 
% will enlarge or reduce the postscript figures based on the xsize
%\epsfbox{azores2.ps}
\end{center} 
% postscript image file name
\caption{The three great leaders trying to convince us that
the weapons of mass destruction exist.  \label{azores}}
\end{figure}

More seriously, 
let us briefly review the history of string phenomenology concerning the
search of the standard model. May be, in this way, we will be able to have a 
more clear opinion about whether the (string) standard model can be found
in the near future. 
In a sense, the compactification
of the ten-dimensional
%$E_8\times E_8$
heterotic string \cite{Gross} on six-dimensional spaces might be
consider
as
the starting point for this race \cite{Candelas}.
% \cite{Candelas}$^-$\cite{Bachas}.
In particular, Calabi-Yau spaces \cite{Candelas},
orbifolds \cite{Dixon} and fermionic constructions \cite{Bachas}
proved to be interesting methods to carry out the 
task. It was shown pretty soon that these compactifications 
of the $E_8\times E_8$ heterotic string can give
rise to four-dimensional standard-{\it like} 
models as well as GUT-{\it like} models \cite{Ross}$^-$\cite{Ellis}.
Clearly, these results were extremely interesting. Since then we know
that (at least) something close to the real world can arise from 
strings.

\begin{figure}[t]
%\figurebox{20pc}{15pc}{azores2.ps} 
%\figurebox{1pc}{1pc}{azores2.ps}
% to have a box alone
\begin{center}
\epsfxsize=12pc 
% will enlarge or reduce the postscript figures based on the xsize
\epsfbox{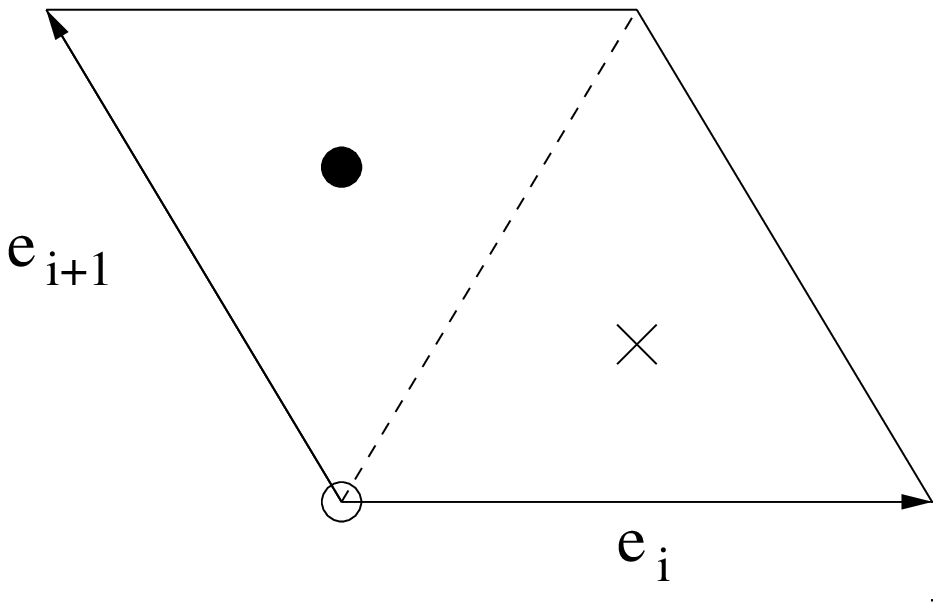}
\end{center} 
% postscript image file name
\caption{Two dimensional sublattices ($i=1,3,5$) 
of the $Z_3$ orbifold. The fixed point components are also shown.
\label{lattice}}
\end{figure}

For the sake of concreteness, let us review the case of
orbifolds, without entering into many mathematical details or
technicalities. 
It was first shown that 
the use of discrete Wilson lines \cite{Dixon,Wilson} on the torus defining a
symmetric orbifold can give rise to four-dimensional
supersymmetric models with gauge group \cite{Kim,Mas}
$SU(3)\times SU(2)\times U(1)^5\times G_{hidden}$.
%and three generations of chiral particles 
%with the correct
%$SU(3)\times SU(2)$ representations 
% (plus extra particles) \cite{Kim,Mas}. 
In addition, it was also shown that three generations
of chiral particles (plus extra particles)
appear in a natural way using just two Wilson lines.
In fact this result was obtained in the case of the $Z_3$ orbifold.
The latter is constructed by dividing 
$R^6$ by the $[SU(3)]^3$ root lattice modded by the
point group ($P$) with generator $\theta$,
where the action of $\theta$ on the lattice basis is
$\theta e_i=e_{i+1}$, 
$\theta e_{i+1}=-(e_i+e_{i+1})$, with $i=1,3,5$.
The two-dimensional sublattices associated to $[SU(3)]^3$ are shown
in Fig.~\ref{lattice}.
In orbifold constructions, twisted strings appear 
attached to fixed points under the point group.
In the case of the $Z_3$ orbifold there are 27 fixed points under
$P$, and therefore there are 27 twisted sectors. 
We will denote the three fixed points of each two-dimensional
sublattice as shown in Fig.~\ref{lattice}. 
Thus the three generations arise 
because in addition to the overall factor of 3 coming from
the right-moving part of the untwisted matter, the twisted matter
come in 9 sets with 3 equivalent sectors on each one. 
Let us suppose that the two Wilson lines
correspond to the first and second sublattices
The three generations correspond to move the third
sublattice component (x $\cdot$ o) of the fixed point keeping the
other two fixed.
%In this way, all matter (including extra
%particles which are usually present) 
%in these constructions appear automatically with three
%generations.

%In this framework it was first possible to obtain
%%It was first shown that the use of background fields 
%% (Wilson lines) \cite{Dixon,Wilson} on the torus defining the
%%symmetric $Z_3$ orbifold can give rise to 
%four-dimensional
%supersymmetric models with gauge group
%$SU(3)\times SU(2)\times U(1)^5\times G_{hidden}$
%and three generations of chiral particles 
%%with the correct
%%$SU(3)\times SU(2)$ representations 
%(plus extra particles) \cite{Kim,Mas}.
%%In fact, it was also shown that the three generations
%%appear in a natural way using just two discrete Wilson lines.
%%This is so because in addition to the overall factor of 3 coming from
%%the right-moving part of the untwisted matter, the twisted matter
%%come in 9 sets with 3 equivalent sectors on each one, since there
%%are 27 fixed points. In this way, all matter (including extra
%%particles) in these constructions appear automatically with three
%%generations.

The next step was the calculation of the $U(1)$ charges
and the study of the mechanism for anomaly cancellation
in these models \cite{Katehou}, since an 
anomalous $U(1)$ is usually present after 
compactification \cite{FayetIliopoulos}.
This allowed the construction of combinations of the non-anomalous $U(1)$'s
giving the physical hypercharge for the particles of the standard
model,
although it was also found that the hidden sector is, in general,
mixed with the observable one through the extra $U(1)$ charges.
Fortunately, it was also noted that the 
Fayet--Iliopoulos D-term \cite{FayetIliopoulos}, 
which appears because of the presence of 
the anomalous $U(1)$, can give
rise to the breaking of the extra $U(1)$'s and,
as a consequence, to the hiding of the previously mixed
hidden sector \cite{Katehou,Casas1}.
This is because, in order to preserve supersymmetry at
high energies, some scalars with $U(1)$'s quantum numbers 
acquire large vacuum expectation values (VEVs).
%It was also noted that the 
%Fayet-Iliopoulos D-term \cite{FayetIliopoulos}, 
%which appears because of the presence of 
%the anomalous $U(1)$, can give
%rise to the breaking of the extra $U(1)$'s.
%This is because, in order to preserve supersymmetry at
%high energies, some scalars with $U(1)$'s quantum numbers 
%acquire large vacuum expectation values (VEVs).
In this way it was possible to construct supersymmetric 
models \cite{Casas1} (or, more precisely,
vacuum states) where the original 
$SU(3)\times SU(2)\times U(1)^5\times SO(10)\times U(1)^3$ 
gauge group \cite{Kim}
was broken to 
$SU(3)\times SU(2)\times U(1)_Y\times SO(10)_{hidden}$. 
In addition, the extra particles are highly reduced since many of
them get a high mass 
($\approx 10^{16-17}$ GeV) through the
Fayet--Iliopoulos mechanism, 
thus disappearing from the low-energy theory.

But...is a model with the gauge group of the standard model
and three families of quark and leptons, the sought-after
standard model?
By no means! For this 
the right model must reproduce also the correct
mass hierarchy for quarks and leptons. For example, to obtain
\begin{equation}
\frac{m_t}{m_u}\sim 10^5\ ,\;\;\;\;\; \frac{m_{\tau}}{m_e}\sim 10^3\ ,
\label{hierar}
\end{equation}
is not a trivial task, although
it is true that
one can find interesting results in the literature concerning this point.
In particular, orbifold spaces have a beautiful mechanism to generate
a mass hierarchy at the renormalizable level.
Namely, Yukawa couplings between twisted matter can be explicitly computed 
%\cite{} 
and they get suppression factors, 
which depend on the
distance between the fixed points to which the relevant fields are
attached \cite{Hamidi}$^-$\cite{Abel}. The couplings 
can be schematically written as
\begin{equation}
\lambda\sim e^{-\sum_i c_{\lambda}^{i} T_i}\ ,\;\;\;\;\; Re\ T_i\sim R_i^2\ ,
\label{supression}
\end{equation}
where the $T_i$ are the moduli fields associated to the size and shape of the
orbifold. The distances can be varied by giving different VEVs to these
moduli, implying that one can span in principle five orders of magnitude the
Yukawa couplings \cite{test,Abel}.

Unfortunately, this is not the end of the story.
As if to obtain
this hierarchy
were not difficult enough, Nature is even more cruel
with string phenomenologists. It tells us that a weak coupling
matrix exists \cite{Cabibbo} with weird magnitudes for the entries
\cite{pdg1}
\bea
%\[
V_{CKM}
=\left( \begin{array}{ccc}
0.9741\ to\ 0.9756 & 0.219\ to\ 0.226 & 0.0025\ to\ 0.0048\\
0.219\ to\ 0.226 & 0.9732\ to\ 0.9748 & 0.038\ to\ 0.044\\
0.004\ to\ 0.014 & 0.037\ to\ 0.044 & 0.9990\ to\ 0.9993
\end{array}\right) 
\ ,
\label{ckm}
\eea
%,\]
%
and that therefore we must arrange our up-and down-quark 
Yukawa couplings in order
to have specific off diagonal elements,
\begin{equation}
H_2^0 \bar u_{L\alpha} \lambda_u^{\beta\gamma} u_{R\gamma}
+
H_1^0 \bar d_{L\alpha} \lambda_d^{\beta\gamma} d_{R\gamma}\ .
\end{equation}
%
%The matrices that diagonalize the 
%$\langle H_2^0\rangle \lambda_u^{\beta\gamma}$ and 
%$\langle H_1^0\rangle \lambda_d^{\beta\gamma}$ 
%mass matrices, 
%should give rise to matrix (\ref{ckm}).
In principle this property arises naturally in 
orbifolds \cite{Ib.}$^-$\cite{Abel}$^,$\cite{oleg}. For example, in the
$Z_3$ orbifold with two Wilson lines,
if the SU(2) doublet $H_2$ corresponds to (o o o), the three generations of 
(3,2) quarks to (o o (o, x, $\cdot$)) and the three generations
of ($\bar{3}$, 1) up-quarks to (o o (o, x, $\cdot$)),
then there are three couplings allowed from the space group
selection rule (the components of the three fixed points
in each sublattice must be either equal or different): 
$\lambda_{tt} H_2^0 \bar t_{L} t_{R}$ associated to 
(o o o)(o o o)(o o o) with $\lambda_{tt}\sim$ 1,
$\lambda_{cu} H_2^0 \bar c_{L} u_{R}$ associated to 
(o o o)(o o x)(o o $\cdot$) with
$\lambda_{cu}\sim e^{-T_5}$, and 
$\lambda_{uc} H_2^0 \bar u_{L} c_{R}$
associated to (o o o)(o o $\cdot$)(o o x) with
$\lambda_{uc}\sim e^{-T_5}$.
In this simple example one gets one diagonal Yukawa coupling
without suppression factor and two off diagonal degenerate ones 
$\sim e^{-T_5}$, but other more realistic examples 
producing the observed structure of quark and lepton masses
and mixing angles
can be obtained
using 
three generations of Higgses \cite{Abel} or three Wilson 
lines \cite{test}.

Unfortunately, although the mechanisms discussed above 
% (\ref{supression},\ref{??}) 
are attractive,
it is extremely difficult to implement them in a particular model.
Given a model, everything is essentially fixed, and it is not possible
to play around. For example, one can have a model with the coupling 
for the bottom allowed
but not for the top, or with both forbidden,  
or with no element 13 in the CKM matrix (\ref{ckm}), or...
The truth is that {\it no model} has been found with all the necessary
Yukawa couplings. As a matter of fact, not even a model close to obtain them.
And this sentence can also be applied to any of the interesting models
constructed in more recent years \cite{Kimm,Cleaver}.

In this sense, in my opinion 
the main difficulty in string model building resides
in how to obtain the weird structure of fermion masses and mixing angles.
My good friend and old collaborator Alberto Casas always says,
`I wouldn't mind to die once I knew the mechanism generating the CKM
matrix'. 
Thus, please, do not find it too soon, we want him to stay
alive,
at least for a while. 

Needless to say, the recent experimental confirmation
of neutrino masses makes this task even more involved. Now, in addition
to hierarchies such as those shown in eq.~(\ref{hierar}), we have to explain
others such as
\begin{equation}
\frac{m_e}{m_{\nu}}\gsim 10^6\ ,
\label{hierarneu}
\end{equation}
and in addition to the matrix (\ref{ckm}), we have to explain
the weak coupling matrix \cite{mns} with
the charged leptons \cite{Torrente}
\bea
%\[
V_{MNS}
=\left( \begin{array}{ccc}
0.73\ to\ 0.89 &\ 0.45\ to\ 0.66 &\ < 0.24\\
0.23\ to\ 0.66 &\ 0.24\ to\ 0.75 &\ 0.52\ to\ 0.87\\
0.06\ to\ 0.57 &\ 0.40\ to\ 0.82 &\ 0.48\ to\ 0.85
\end{array}\right) 
\ .
\eea
Again, as usual in string theory, one can find interesting mechanisms
to try to explain these experimental results.
For example,
if the Yukawa coupling
for the neutrino is of order \( m_{e} \) and the see saw scale is
1 TeV, then the expected neutrino mass is
\bea
\frac{m_{e}^{2}}{1\ TeV}=0.25\ eV\ ,
\label{seesawmass}
\eea
which is within an order of magnitude of the experimental values.
This suggests that a natural situation is one in which a see-saw mass 
of order a few TeV is generated by the electroweak symmetry breaking. 
The first guess for the neutrino see-saw superpotential is then \cite{Abel}
\bea
W_{\nu}\sim  \lambda_{\nu} H_2^0 L_L\nu_R + \lambda_N N\nu_R \nu_R
%+ NH_2^0H_1^0
\ ,
\label{seesaw}
\eea
where \( N \) is the same singlet that dynamically generates the
$\mu$ term through the coupling \footnote{Note 
in this context that the Giudice--Masiero 
mechanism to generate a $\mu$ term through the
K\"ahler potential is not available for prime orbifolds such as the
$Z_3$ orbifold.}
%and that this orbifold, where three generations appear authomatically
%using two discrete Wilson lines, is the one where the
%above mentioned standard-like models were constructed.
$NH_2^0H_1^0$ . 
Therefore $N$ is expected to get a VEV of order 1 TeV, and the coupling
$\lambda_{\nu}$ is expected to be sufficiently small as to reproduce the
neutrino mass. Since small couplings can be naturally
obtained in orbifolds as discussed in eq.~(\ref{supression}), 
this mechanism is in principle
interesting.
But recall: to implement any mechanism in a particular model is 
highly non-trivial.

However, since we are optimistic people, we can argue that if the
standard model arises from strings (something that we believe 
firmly!) there must exist one model where the above mechanism
can be implemented, producing the
{\it correct structure} for Yukawas. This means a model with:
1) the necessary Yukawas couplings, 
$H_2^0 \bar u_{L} \lambda_{uu} u_{R}
+
H_2^0 \bar u_{L} \lambda_{uc} c_{R}
+
H_1^0 \bar d_{L} \lambda_{db} b_{R}
+...$,
2) the correct values,
i.e. at least one is able to put by hand the values of 
$T_i$ such that $\lambda_t(T_i)\sim 1$,
$\lambda_u(T_i)\sim 10^{-5}$, etc.
If, at the end of the day, such a model exists this would be 
a great success.
But, in order to be sure that this is really 
the superstring standard model one should be able to 
compute explicitly the values of the Yukawas, and for this
we need to know
the VEVs of the $T_i$-moduli. Unfortunately, these are related
to the breaking of supersymmetry, and this is one of the 
biggest problems in string theory\footnote{ 
As a matter of fact we should
also be able to compute the values of the gauge couplings
$g_3,g_2,g_1$,
determined by the VEV of the dilaton field $S$. Let us recall
that this field arises
from the gravitational sector of the theory, and that in 
string theory there are no free parameters, all
coupling constants are in fact no constants but fields.}.
It is true that there are candidates for this task, such as gaugino
condensation in a hidden sector with a non-perturbative superpotential
$W(S,T_i)$, and that we have hidden gauge groups
that could condensate. However, again, implementing this mechanism
in a particular model is not easy.

\begin{figure}[t]
%\figurebox{20pc}{15pc}{azores2.ps} 
%\figurebox{1pc}{1pc}{azores2.ps}
% to have a box alone
\begin{center}
\epsfxsize=13.5pc 
% will enlarge or reduce the postscript figures based on the xsize
%\epsfbox{puzzle.ps}
\hspace{0.1cm}
\epsfxsize=13.7pc 
%\epsfbox{puzzle2.ps}
\end{center} 
% postscript image file name
\caption{Two transparencies summarizing 
the situation concerning the search of the
standard model in string phenomenology in 1990 (still can be
used in 2003!). \label{trans}}
\end{figure}

The above discussion could be summarized with the two transparencies
shown in Fig.~\ref{trans}.
What is annoying for me is that I prepared these two transparencies 
for a meeting in Trieste \cite{trieste} in 1990, 
and still I can use them in this
meeting 13 years later! In a sense the problem
of string theory is that it is too ambitious: the correct model
must reproduce not only the gauge group and families of the standard model,
but also the correct values of the gauge couplings,
the correct masses of quark and leptons, a realistic CKM matrix, etc., 
i.e. the more than 20 parameters fixed
by the experiment in the standard model. 

In addition, there are thousands of models 
(vacua) that can be built.
Some of them have
the gauge group of the standard model or GUT groups,
three families of particles, and other interesting properties, but
many others have 
a number of families different from three, 
no appropriate gauge groups, no appropriate matter, etc.
A perfect way of solving this problem would be to use
a dynamical mechanism to select the correct model (vacuum).
Such a mechanism should be able to determine a 
point in the parameter space of the heterotic string
determining the correct
compactification 
with
$SU(3)\times SU(2)_L\times U(1)_Y$, three families of particles,
and such that the mechanism of supersymmetry breaking (whatever it is)
produces $\langle S,T_i\rangle$ generating
the correct values for 
\begin{equation}
g_3,g_2,g_1,\lambda_u,\lambda_d,\lambda_c,
\lambda_s,\lambda_t,\lambda_b,\lambda_e,\lambda_{\mu},\lambda_{\tau},
\lambda_{\nu_e},
\lambda_{\nu_{\mu}},\lambda_{\nu_{\tau}},\delta,\theta_c,...
\end{equation}
In a sense, it is hard to believe that there exists a (top-bottom) mechanism
with such a precision determining everything. 
But here we apply again our optimism, arguing that the standard
model must arise from strings, and that therefore such a marvellous
mechanism must exist.
The only problem is that...it has not been discovered yet.

So, for the moment, the best we can do is...keep trying!,
i.e. to use the experimental results available (such as 
$SU(3)\times SU(2)_L\times U(1)_Y$, three families,
fermion masses, mixing angles, etc.), 
to discard models.
Although the model space is in principle huge, a detailed analysis
can reduced this to a reasonable size.
For example, within the 
$Z_3$ orbifold with two Wilson lines,
one can construct in principle a number 
of order 50000 of three-generation 
models
with the
$SU(3)\times SU(2)\times U(1)^5$ gauge group 
%in the observable sector
associated to the first $E_8$ of the heterotic string.
However, a study implied that most of them are equivalent
\cite{Mondragon}, 
and in fact, at the end of the day, only 192 different models were
found \cite{Giedt,Mondragon}.
This reduction is remarkable, but we should keep in mind that
the analysis of each one of these models is painful. 

In summary, to obtain a 
connection between (string) theory and present (standard model)
experiments is possible in principle
but difficult in practice. 
But, what about future experiments (such as LHC)?
Well,
if Nature is supersymmetric at the weak scale, 
as many particle physicists
believe (ironically string phenomenologists, at least some of
them, who were originally supersymmetry phenomenologists, 
are not so enthusiastic nowadays
with supersymmetry as in the past because of the recent developments), 
eventually the spectrum of supersymmetric
particles will be measured providing us with a possible connection 
with the high--energy world of 
superstrings.
Let us recall that in superstring constructions there is
a natural hidden sector built in, the singlet fields $S$ and $T_i$
mentioned above,
and that the K\"ahler potential 
$K(S,S^*,T_i,T_i^*)$ and the gauge kinetic function
$f(S,T_i)$ of the four-dimensional supergravity Lagrangian are known.
As a consequence, the soft supersymmetry-breaking terms,
scalar masses $m_{\alpha}$, gaugino masses $M_a$, etc.,
can be computed in principle and compared with the experimentally observed
supersymmetric spectrum \cite{brignole}, i.e. we will be able to
do what one could call \cite{soft} ``Soft Phenomenology''.

Although this approach will not probably be sufficient to select
the superstring standard model, at least it will allow us to
discard many constructions not producing the correct values for
the soft terms. In addition, if experimentalists find some
extra particles which arise naturally in a particular string
framework \cite{kind}, 
this might be helpful.

In any case we should not forget that the cosmological
constant contributes to the value of the soft terms,
introducing in principle another problem in our discussion.
First of all, we have a new free parameter in the computation,
e.g. $m_{\alpha}^2 \sim m^2_{3/2} + V_0/M_P^2$.
Second, as is well known, in any theory including gravity
the natural value of the cosmological constant is huge
(in our case 
$V_0\sim m^2_{3/2} M_P^2$ once supersymmetry is broken),
and this
is one of the biggest problems in particle physics.
If we use a specific mechanism for the breaking of supersymmetry,
all this may be specially disturbing, e.g. in gaugino condensation
$V_0$ turns out to be negative and including this contribution
one might obtain $m_{\alpha}^2<0$.
\begin{figure}[t]
%\figurebox{20pc}{15pc}{azores2.ps} 
%\figurebox{1pc}{1pc}{azores2.ps}
% to have a box alone
\begin{center}
\epsfxsize=12pc 
% will enlarge or reduce the postscript figures based on the xsize
%\epsfbox{pandoraletras.ps}
%\hspace{1cm}
%\epsfxsize=10pc 
%\epsfbox{pandora.ps}
\end{center} 
% postscript image file name
\caption{Curious scientist opening the Padora's box 
of D-brane constructions. Multitude of `plagues' 
for hapless string phenomenologists escape. 
\label{pandora}}
\end{figure}

\vspace{0.5cm}

\noindent Until recently, the 
$E_8\times E_8$ heterotic superstring framework discussed above
was thought as the only way in order to construct realistic
string models. However,
in the late nineties it has been
discovered that 
D-brane configurations from 
%type I 
string vacua 
%\cite{dbranes}
or heterotic M-theory 
%\cite{mtheory}
can also give rise to 
explicit models, with interesting phenomenological 
properties \cite{dmodels,ovrut} (although with unrealistic Yukawas
for the moment, as in the case of the perturbative heterotic models).
%can also be constructed using D-brane configurations from 
%type I string vacua \cite{dbranes}
%or heterotic M-theory \cite{ovrut}.
Of course, this means that we have more work to do since we have
more models to analyze, but also...
that the Pandora's box opened.
As you know, Jupiter had crammed 
into a box all the diseases, sorrows, vices that afflict poor
humanity. Pandora, the first woman, who did not know this, 
was seized with an eager curiosity to know what the box contained.
One day she slipped off the cover and...forthwith there escaped
all plagues for hapless man 
In our case, as shown in Fig.~\ref{pandora},
when D-branes are included in the game,
many possibilities completely forbidden in the
context of the heterotic string
are now allowed: Non-supersymmetric models can be constructed \cite{nonsusy}, 
the string scale 
$M_{string}$ may be
anywhere between the weak scale $M_W$ and the Planck scale
$M_{Planck}$ \cite{weakplanck}, 
large extra dimensions are possible \cite{extradi}, etc.
Exaggerating, now the question is not the old one: 
What is possible to predict in string theory?,
but,
What {\it is not} possible to predict?

Let me point out, however, 
that some of these possibilities imply a hierarchy problem.
For example, embedding the standard model inside D3-branes, one has 
%the string scale is given by:
%
\bea
\frac{M_{string}^4}{M_c^3}= \frac{\alpha M_{Planck}}{\sqrt 2}=
3.5\times 10^{17}\ GeV\ ,
\label{gravitino}
\eea
where $\alpha$ is the gauge coupling and $M_c$ is the compactification
scale
in the case of an overall modulus $T$. 
Thus one gets $M_{string}\approx 10^{11}$ GeV much smaller than the
Planck scale 
with $M_c\approx 10^{9}$ GeV, i.e. an apparently modest input hierarchy.
However, in fact those values would imply 
\bea
S=\frac{1}{\alpha} \simeq 24\ ,\,\,\,\,\,
T=\frac{1}{\alpha}\left(\frac{M_{string}}{M_c}\right)^4\simeq 10^9\ .
\label{gravitino2}
\eea
Thus one has a hierarchy problem but with the VEVs of the
fields that one has to determine dynamically.
Of course, if we want to lower 
$M_{string}$ further the hierarchy problem is worse.
E.g. using eq.~(\ref{gravitino}) for the case of two different
compact dimensions $M_1\sim 10^{-13}$ GeV (i.e. 1 millimeter)
and $M_2= M_3 \sim 10^4$ GeV, one obtains
$M_{string}= (M_1 M_2^2\times 3.5\times 10^{17}\ GeV)^{1/4}\sim 1$ TeV.
But then,
\bea
S=\frac{1}{\alpha} \simeq 24\ ,
T_2=\frac{1}{\alpha}\left(\frac{M_{string}^4}{M_1^2M_3^2}\right)\simeq 
10^{31}\ ,
T_1=\frac{1}{\alpha}\left(\frac{M_{string}^4}{M_2^2M_3^2}\right)\simeq 
10^{-3}\ .
\label{gravitino2}
\eea

\begin{table}[t]
\caption{Statistics of the meeting String Phenomenology 2003,
concerning the subjects of the talks. \label{statistics}}
\begin{center}
\footnotesize
\begin{tabular}{|c|c|c|l|}
\hline
{\bf Subjects} &\raisebox{0pt}[13pt][7pt]
{\bf Number of talks}\\
\hline
%\cline{1-2}
{Branes} &   {17}
\\ 
\hline
{Heterotic string} & {8}
\\ 
\hline
{M-theory} & {7}
\\ 
\hline
{Supersymmetry} & {7}
\\ 
\hline
{Cosmology} & {5}
\\ 
\hline
{D=5 constructions} & {3}
\\ 
\hline
{Accelerators} & {2}
\\ 
\hline
{Quantum gravity} & {1}
\\ 
\hline
{AdS} & {1}
\\ 
\hline
{Non-commutative} & {1}
\\ 
\hline
{Electroweak} & {1}
\\ 
\hline
{Neutrinos} & {1}
\\ 
\hline
{...} & {...}
\\ 
\hline
\end{tabular}
\end{center}
\end{table}

In any case, it is clear that nowadays 
these constructions are the new super-stars, and that
supersymmetry, Planck scale physics, small extra dimensions, 
or anything not involving D-branes is...out of fashion!
Please, do not feel attack if you still work in these out-of-fashion
issues...
Of course I am joking and exaggerating,
and in fact after following the talks of this meeting my impression is that
(although clearly the winners are the Branes)
the `old' constructions, such as the perturbative heterotic string,
are having a revival. See Table~\ref{statistics} for more details.

\begin{figure}[t]
%\figurebox{20pc}{15pc}{azores2.ps} 
%\figurebox{1pc}{1pc}{azores2.ps}
% to have a box alone
\begin{center}
\epsfxsize=17pc 
% will enlarge or reduce the postscript figures based on the xsize
%\epsfbox{evolution.ps}
\end{center} 
% postscript image file name
\caption{Evolution: from the monkey to the string phenomenologist.  
\label{evolution}}
\end{figure}

Finally, given the previous discussion, 
let me summarize the evolution of string phenomenology
using a few figures.
First of all,  
the evolution from the monkey to the string phenomenologist is shown
in Fig.~\ref{evolution}.
As is well known, the string phenomenologist is a fellow 
who spends most of the time using her/his computer
compulsively: Sending revised versions to hep-ph
(and even to hep-th), complaining about references, preparing
talks...
Fortunately, from time to time she/he has a small hole in 
her/his tight schedule and spend some time thinking
about string phenomenology.
This is precisely the moment shown in Fig.~\ref{cuerdas2}.
Clearly, the poor guy sit down at the front of the room
is a string phenomenologist. In 1985 he really believed that the standard model
was somewhere around him and that he would be able to find it.
The problem, as you can see, is that the search of the standard model 
in strings
is like to look a needle in a haystack.
What about the guy at the back of the office?
Well, he is clearly a string theoretician. As you can see he
seems quite confident having a look to his very nice
classification of string theories, $E_8\times E_8$ heterotic, 
type I, type IIA, ..., and with all the machinery to work
with them in the shelves. But what for this theoretician is
a nice classification, for the phenomenologist is a 
horrible nightmare. He has a incredible mess, a jungle, in his hands. 
As shown in 
Fig.~\ref{cuerdas3}, because of the huge number of models, each one with 
its own characteristics, the situation was, in a sense, depressing.

\begin{figure}[t]
%\figurebox{20pc}{15pc}{azores2.ps} 
%\figurebox{1pc}{1pc}{azores2.ps}
% to have a box alone
\begin{center}
\epsfxsize=28pc 
% will enlarge or reduce the postscript figures based on the xsize
%\epsfbox{cuerdas.ps}
\end{center} 
% postscript image file name
\caption{A string phenomenologist searching the standard model
in 1985.  \label{cuerdas2}}
\end{figure}
%
%\vspace{-3cm}
%
\begin{figure}[t]
%\figurebox{20pc}{15pc}{azores2.ps} 
%\figurebox{1pc}{1pc}{azores2.ps}
% to have a box alone
\begin{center}
\epsfxsize=24pc 
% will enlarge or reduce the postscript figures based on the xsize
%\epsfbox{cuerdasconflechas2.ps}
\end{center} 
% postscript image file name
\caption{The huge number of possibilities in the model space
that a string phenomenologist found in the analysis in 1985. 
\label{cuerdas3}}
\end{figure}

%\newpage

Now, we can see the evolution in the search of the standard model
comparing Figs.~\ref{cuerdas2} and  \ref{cuerdas4}.
In the latter we see the same string phenomenologist but in 2003.
Do you see any difference? Clearly, the mess is exactly the same: 
terrible\footnote{An optimistic view of this situation was pointed out
to me by the experimentalist H.U. Martyn. In his opinion,
working in string phenomenology is healthy because
18 years have gone by between the two photographs
and the guy does not look old at all!}.
Wait a moment, 
this is not true, as shown in Fig.~\ref{cuerdas5} the mess is even worse!
Because of the recent developments,
the number of possible models is now bigger, and therefore the number
of different results increases.

Anyway, the meeting is finishing, you are going 
back home, and I do not want you to leave Durham 
crying and depressed. So, let me tell you something very optimistic
about string phenomenology.
Not only the topic:
from another viewpoint this situation is good because this
means that still there is a lot work to do for newcomers.
I am in the position of telling you something much stronger.
Usually people says that the problem of strings is that they do not
have a clear prediction that can be tested. On the contrary, I can tell
you very proudly that predictions can be done. Indeed, I did 
an important prediction using strings two years ago.
This prediction has been finally fulfilled. The British experimentalists
have tested it very recently.  
Please, have a look to Fig.~\ref{prediction}, where the abstract
and date of the article containing the prediction is shown.
Thank you very much for your attention.
\begin{figure}[t]
%\figurebox{20pc}{15pc}{azores2.ps} 
%\figurebox{1pc}{1pc}{azores2.ps}
% to have a box alone
\begin{center}
\epsfxsize=16.7pc 
% will enlarge or reduce the postscript figures based on the xsize
%\epsfbox{cuerdas.ps}
\end{center} 
% postscript image file name
\caption{Evolution of string phenomenology:
The same string phenomenologist searching the
standard model as in Fig.~\ref{cuerdas2},
{\it but in 2003}. Do you see any evolution?
\label{cuerdas4}}
\end{figure}

\begin{figure}[t]
%\figurebox{20pc}{15pc}{azores2.ps} 
%\figurebox{1pc}{1pc}{azores2.ps}
% to have a box alone
\begin{center}
\epsfxsize=24pc 
% will enlarge or reduce the postscript figures based on the xsize
%\epsfbox{cuerdasconflechas.ps}
\end{center} 
% postscript image file name
\caption{Because of the recent developments,
the number of possibilities in 2003 is now bigger than
in 1985 (see Fig~\ref{cuerdas3}):
the nightmare is even worse! 
\label{cuerdas5}}
\end{figure}

%\end{document}

\begin{figure}[t]
%\figurebox{20pc}{15pc}{azores2.ps} 
%\figurebox{1pc}{1pc}{azores2.ps}
% to have a box alone
\vspace{-2cm}
\begin{center}
\epsfxsize=25pc 
% will enlarge or reduce the postscript figures based on the xsize
\epsfbox{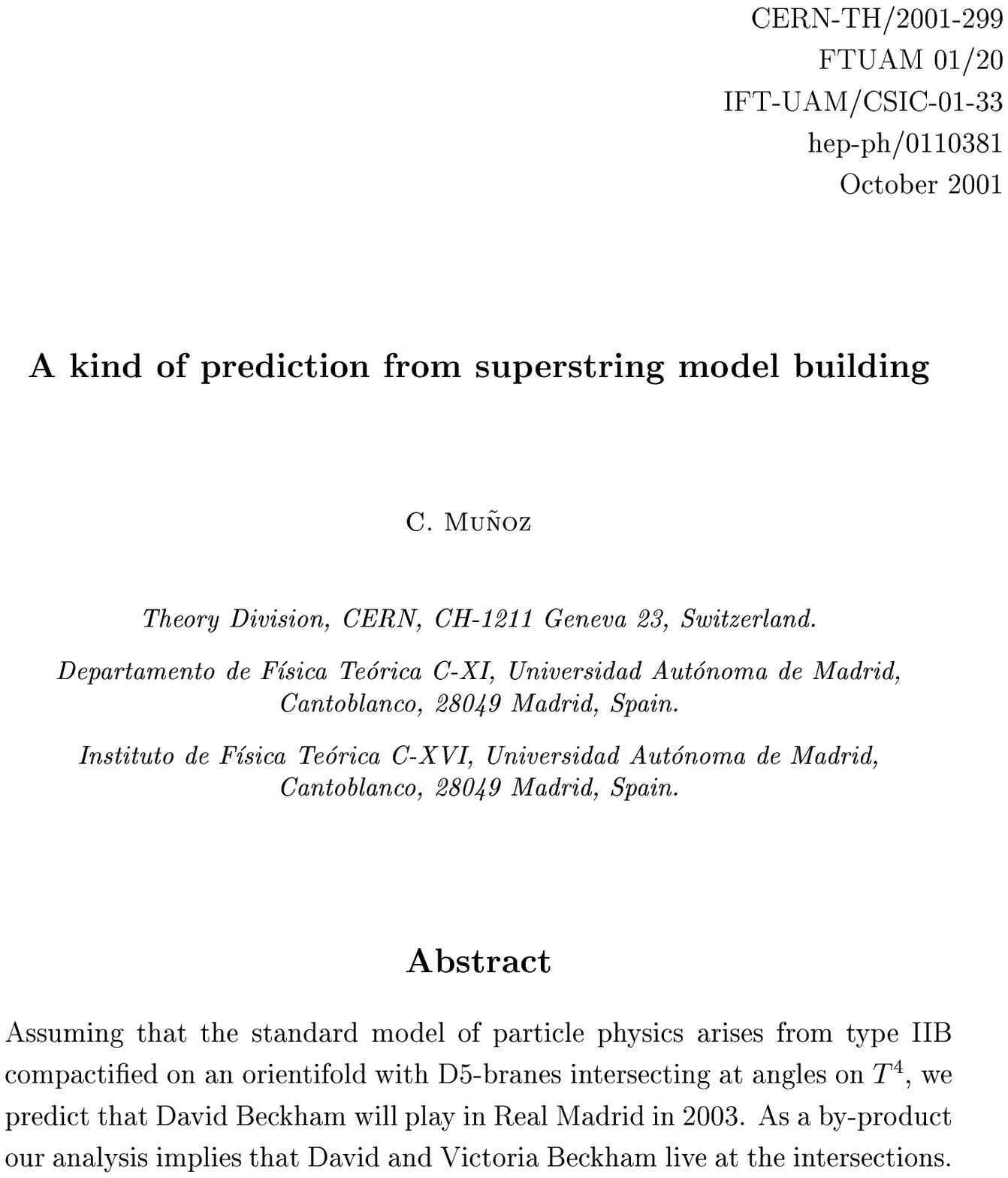}
\end{center}
\vspace{-4cm} 
% postscript image file name
\caption{First page of an article containing a clear prediction
from strings, that has been fulfilled.  \label{prediction}}
\end{figure}

\section*{Acknowledgments}

I thank the organizers of this wonderful conference in Durham 
during July 29-August 4, 2003, for
suggesting me the idea of giving an
%pushing me to give an 
`entertaining' talk about 
string phenomenology. At least I tried.

The photograph shown in Figs.~6 and 8 is a work of art made
by Jeff Wall in 1994, with the title ``Untangling''. Although 
several colleagues thought that I was the author, I only 
deserve the reputation of the modifications shown
in Figs.~7 and 9.
When I first saw this photograph in October 1999, in an exhibition
of the Pamela and Richard Kramlich collection of media art
at San Francisco Museum of Modern Art, I immediately thought
`This is the best possible summary of the state-of-the-art in
string phenomenology'.

%\section*{Appendix}
%We can insert an appendix here and place equations so that they
%are given numbers such as Eq.~(\ref{eq:app}).
%\be
%x = y.
%\label{eq:app}
%\ee

%\newpage

\end{document}